# Benchmarking of hydrodynamic plasma waveguides for multi-GeV laser-driven electron acceleration

B. Miao[1], E. Rockafellow[1], J.E. Shrock[1], S.W. Hancock[1], D. Gordon[2] and H.M. Milchberg[1,3]

[1]*Institute of Research in Electronics and Applied Physics and Dept. of Physics, University of Maryland, College Park, MD 20742*
[2]*Naval Research Laboratory, Washington DC 20375*
[3]*Dept. of Electrical and Computer Engineering, University of Maryland, College Park, MD 20742*

Hydrodynamic plasma waveguides initiated by optical field ionization (OFI) have recently become a key component of multi-GeV laser wakefield accelerators. Here, we present the most complete and accurate experimental and simulation-based characterization to date, applicable both to current multi-GeV experiments and future 100 GeV-scale laser plasma accelerators. Crucial to the simulations is the correct modeling of intense Bessel beam interaction with meter-scale gas targets, the results of which are used as initial conditions for hydrodynamic simulations. The simulations are in good agreement with our experiments measuring evolving plasma and neutral hydrogen density profiles using two-color short pulse interferometry, enabling realistic determination of the guided mode structure for application to laser-driven plasma accelerator design.

## I. Introduction

Laser wakefield acceleration (LWFA) in plasmas is a promising technique for achieving multi-GeV/m acceleration gradients [1,2]. LWFA has been applied to high energy photon sources [3,4], free-electron lasers [5] and secondary particle generation [6,7]. Staging of multiple LWFAs has been proposed to realize a TeV electron collider for high energy physics [8,9]. Although multi-GeV [10–13], high-charge (~nC) [14], and low energy spread (<1%) [15] electron beams have been demonstrated separately, a laser wakefield accelerator realizing all of these features will be competitive with conventional accelerators provided there is improvement in high-repetition rate, high energy ultrafast lasers [16] and precise control of laser propagation in plasma [17].

Laser wakefield acceleration of electrons to the $\sim 10$ GeV scale in a single acceleration stage requires maintaining a normalized laser vector potential $a_0 > 1$ over tens of centimeters of plasma at low on-axis plasma densities $N_{e0} \sim 10^{17} \text{cm}^{-3}$ [11], demanding some form of laser guiding. Low plasma density imposes steep laser requirements on relativistic self-guiding: petawatt laser power and $a_0 \gg 1$. By contrast, acceleration driven by pulses guided in preformed plasma waveguides requires $a_0 > \sim 1$ and reduced powers of a few hundred TW, a significantly more efficient use of laser energy [11].

The most often-used preformed plasma waveguides have been generated from capillary discharges [10,18,19] or hydrodynamic expansion of elongated laser-induced plasmas [20–35], pioneered in [20, 21]. For future practical laser-driven accelerators at $\sim 10$ GeV and higher, plasma waveguides from laser ionization of meter-scale gas jet sources [11,17, 26-28] are preferable for their wide tunability in refractive index structure and length, their ability to operate at very high repetition rate, and their immunity to laser-induced damage. We have developed two approaches to generate meter-scale, low density plasma waveguides using optical field ionization (OFI) by short pulse Bessel beams [28]: the "2-Bessel" method [26] and the "self-waveguiding" method [27], also shown for high density 1 mm plasmas [29]. These methods are essential to ionize the cylindrical shock in the neutral gas surrounding the OFI-heated plasma to form the waveguide cladding; the initial OFI plasma itself does not form a waveguide structure, as shown in [26-28]. Subsequently, we implemented the self-waveguiding method to demonstrate the first multi-GeV laser wakefield acceleration in an all-optical LWFA experiment [11].

Crucial to further development of low-density meter-scale waveguide-based LFWA is the detailed characterization of the waveguides themselves. In our earlier work [11,27,28], we first demonstrated the use of two-colour interferometry and fluorescence imaging to characterize OFI-induced plasma waveguides, where knowledge of both the plasma and neutral density profiles is essential for implementing the 2-Bessel and self-waveguiding methods. Our more recent multi-GeV acceleration results [17] have highlighted the need for even higher fidelity measurements and simulations; those experiments reveal a new modal propagation effect that depends sensitively on the waveguide structure and, in turn, determines the electron spectra.

Recent work [35] has compared hydrodynamic simulations of OFI-induced plasma waveguide generation to single colour interferometric measurements of the total phase shift profile generated by a lens-focused high intensity pulse [30], assumed in the simulations to have a Gaussian transverse profile unaffected by propagation in the ionizing gas. However, as we will show, generation of meter-scale plasma waveguides necessitates self-consistent propagation simulations to correctly determine the 3D electron density and temperature profiles for use in hydrocode simulations. Furthermore, measurement of both the electron and neutral gas density profiles, lacking in [35], are crucial to proper benchmarking of simulations and to methods for generating the plasma waveguide cladding [26,27].

In this paper, we present the first comprehensive experimental and simulation study of meter-scale plasma waveguides that considers the self-consistent Bessel-beam-induced OFI and heating of a long gas target and the hydrodynamic response of both the plasma and neutral species. We use two-colour interferometry to extract the plasma and neutral hydrogen transverse profile evolution over a 10 ns range, delays sufficiently long to accommodate up to ~100 μm waveguide mode sizes for proposed ~100 GeV accelerator stages. Specifically, we perform simulations [27,36] of (a) self-consistent propagation of intense femtosecond Bessel beams in hydrogen gas for generation and heating of meter-scale plasmas, (b) hydrodynamic evolution [37] of the Bessel beam-initiated plasma and neutral hydrogen profiles, and (c) quasi-bound modal analysis [38] of the simulated waveguides. We compare the hydrodynamic simulations to experiments measuring both the electron density and neutral species profiles.

## II. Experimental setup

Figure 1 shows the experimental setup. A waveguide-forming pulse (λ=800nm, $\tau = $ 50 fs FWHM, energy 150 mJ or 300 mJ, $R = 2$ cm beam radius) was phase corrected by a deformable mirror [39] and focused by a diffractive axicon (fused silica substrate, 0.5 mm thick, 55% first order efficiency) to form a zeroth order Bessel beam pulse (denoted J$_0$) in an experimental chamber with 67 mbar hydrogen backfill (molecular density $1.7 \times 10^{18}$ cm$^{-3}$). The Bessel beam rays approach the optical axis at an angle $\gamma = 2.3°$, giving a geometric focal length [21] of $L_{foc} = R/\tan\gamma \approx$ 50 cm. The phase correction was performed to preserve the J$_0$ profile over the full focal length [39]. Both linearly polarized (LP) and circularly polarized (CP) pulses were used. The ellipticity of the CP pulses was measured to be 0.82 by a camera placed at $z = 20$ cm on the focal line (Fig. 1(a)), where $z = 0$ is the location of the diffractive axicon.

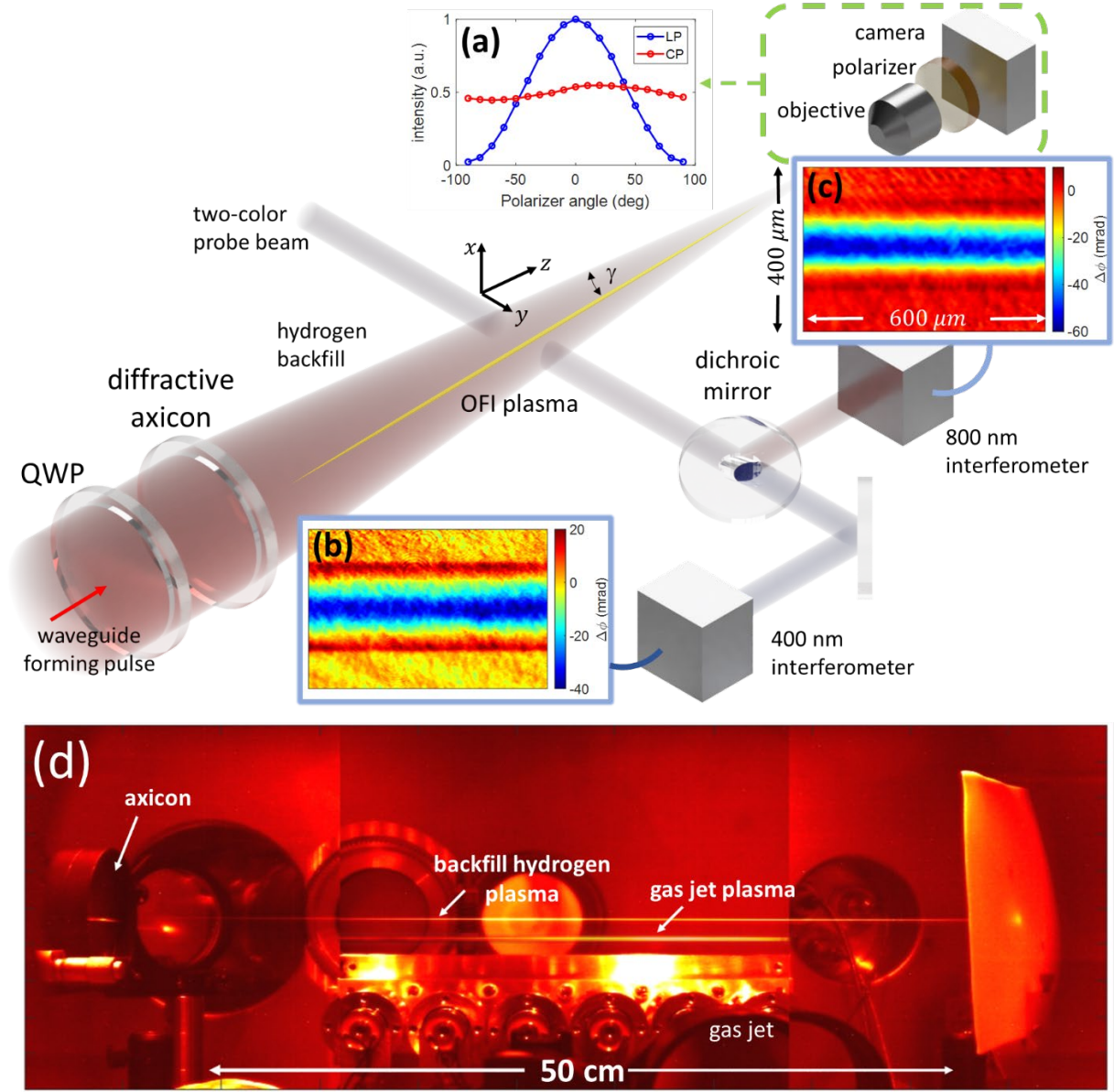

**Figure 1.** Experimental setup showing plasma generation and transverse 2-colour interferometric probing. QWP: λ/4 plate. (a) Measurement of linear polarization (LP, blue) and circular polarization (CP, red), of the Bessel beam. (b) Phase shift image at $\Delta t = 10$ ns with 400 nm probe beam. (c) Phase shift image at $\Delta t = 10$ ns with 800 nm probe beam. (d) Image of 50 cm long hydrogen backfill fluorescence for Bessel beam energy 150 mJ. An image of fluorescence from Bessel-beam-heated 20-cm-long H$_2$ jet plasma is superimposed. The in-situ 20 cm jet, not used in these experiments, is shown for scale.

Two-colour interferometry was used to separately extract the time evolution of the electron and neutral hydrogen density profiles generated by the Bessel beam [27]; a photo of the ~50 cm long hydrogen backfill plasma fluorescence is shown in Fig. 1(d), with the in-situ gas jet shown for scale (along with the superimposed image of 20 cm jet plasma). Auxiliary pulses were split from the main beam before compression, spatially filtered and frequency doubled with a 1-mm thick BBO crystal to form a two-color (400 nm, 800 nm) probe beam, which passed through a $\Delta t = 0\text{–}10$ ns delay line and was then directed transversely through the plasma column at $z = 23$ cm, near the axial centre of the geometric focus. After the plasma, a dichroic splitter directed each colour probe to its own shearing wavefront interferometer, where the phase shift imparted to each probe beam was measured. Phase shift images at 400 nm and 800 nm are shown in Figs. 1(b) and 1(c). To extract the refractive index profiles (and thus the electron and neutral profiles), the phase shift was averaged over 500 shots (frames) as well as over 2500-pixel columns ($\Delta z = 0.78$ mm) along the Bessel beam axis to reduce measurement noise to $< 0.5$ mrad RMS. The spatial resolution of the imaging system is ~3 μm and ~6 μm respectively for the 400 and 800 nm interferometry arms. The plasma and neutral density profiles are extracted from the phase shift profiles by Abel inversion [40]. Further details of interferogram processing can be found in Appendix A.

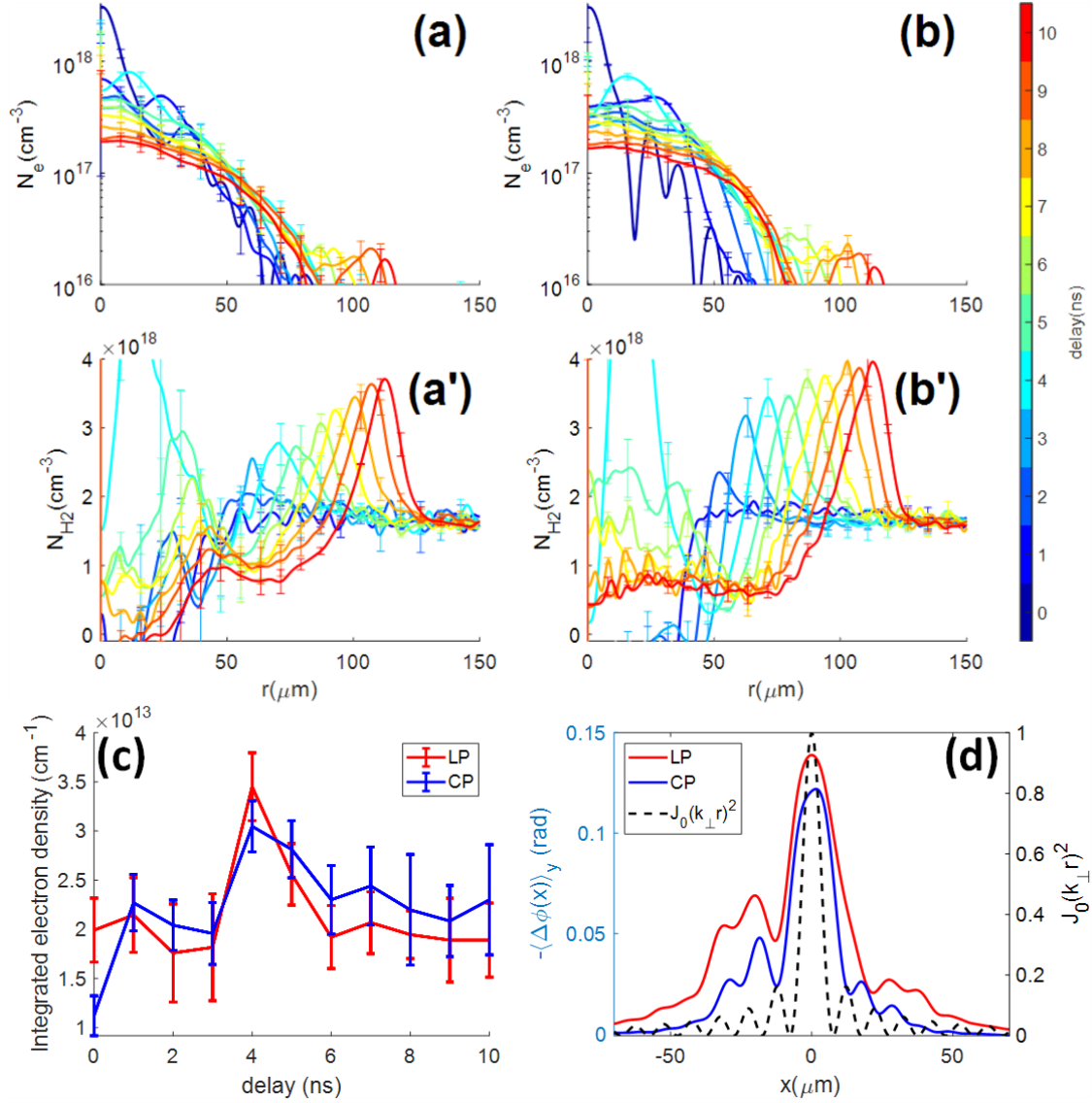

**Figure 2.** Evolution of electron ((a) and (b)) and neutral hydrogen ((a') and (b')) density profiles for delays $\Delta t = 0 - 10$ ns for $\varepsilon_{Bessel} = 150$ mJ, pulsewidth 50 fs, Bessel beam axis approach angle $\gamma = 2.3°$ and beam radius 2 cm. (a) Electron density profiles $N_e(r, \Delta t)$ for LP. (b) Electron density profiles $N_e(r, \Delta t)$ for CP. (a') neutral hydrogen density for LP, (b') neutral hydrogen density for CP, (c) Electron density per unit length $Q(\Delta t) = \int dr 2\pi r N_e(r, \Delta t)$ (d) Plasma phase shift at delay $\Delta t \sim 1$ ps, induced by LP (red) or CP (blue) Bessel beam, measured by the 400 nm probe beam. An ideal Bessel beam profile in vacuum (black) is plotted for reference. The measurement uncertainties are described in Appendix A.

We measured the Bessel beam-induced plasma expansion for both LP and CP J$_0$ pulses in 67 mbar (50 torr) hydrogen backfill for pulse energies $\varepsilon_{Bessel} = 150$ and 300 mJ. The extracted electron and neutral density profiles for $\varepsilon_{Bessel} = 150$ mJ are plotted in Fig. 2(a) and 2(a') for LP and in Fig. 2(b) and 2(b') for CP. The $\varepsilon_{Bessel} = 300$ mJ case is presented in Appendix A. The initial electron density column at $\Delta t \sim 0$ formed by field ionization and heating by the Bessel beam (Fig. 2(a) and 2(b)) shows radial modulations from the Bessel rings. The tranverse electron and ion pressure gradients then drive hydrodynamic expansion of the ionized hydrogen gas, with the on-axis electron density rapidly decreasing and a shock forming between the expanding plasma and the neutral gas on the periphery (Fig. 2(a') and 2(b')). Note that at no point in the hydrodynamic evolution does there develop a concave electron density profile capable of optical guiding, making necessary the auxiliary 2-Bessel or self-waveguiding methods for generating the plasma cladding. These measurements agree with those in [27], which also used Bessel beam ionization and heating, and are in contrast with [35], where a short focal length lens was used for plasma generation, a setup inconsistent with the requirements for meter-scale plasma waveguides.

The shock propagates as cylindrical blast wave expansion at approximately the ion acoustic speed [41], with the CP-induced expansion $\sim 5 - 10\%$ greater than the LP expansion. Interestingly, the neutral density profiles in Fig. 2(a') and (b') peak off centre at $\Delta t \sim 4 - 5$ ns, caused either by phase extraction error for the neutral contribution (in regions where the phase shift is dominated by electrons) or, more likely, the cylindrical implosion of an initial plasma annulus outside the central plasma [42], here driven by the side lobes of the Bessel beam. The neutrals likely originate from un-ionized gas from the Bessel beam nulls that is compressed inward. The peak then relaxes within $\sim 1 - 2$ ns.

While the initial plasma is generated by ultrafast OFI on the few femtosecond timescale, the ionization level of the plasma can later change on the nanosecond hydrodynamic timescale via recombination and collisional ionization. To assess the change in total charge during hydrodynamic expansion, we plot in Fig. 2(c) the radial integral of the electron density profile extracted from Abel inversion of the phase shift measurement, $Q(\Delta t) = \int dr 2\pi r N_e(r, \Delta t)$, which is the electron density per unit length of the plasma channel. The bump in $Q(\Delta t)$ near $\Delta t \sim 4 - 5$ ns is associated with the peaks in Figs. 2(a') and 2(b') at the same delay: the electron density temporarily follows the locally compressed density of heavy particles (see Figs. 2(a) and 2(b)), but the radial wings are below the interferometric density sensitivity. This gives the apparent (but not real) effect of an increase in linear electron density at $\Delta t \sim 4 - 5$ ns. To within error, the charge remains relatively constant for both LP and CP, with the CP result slightly larger at most delays.

Figure 2(d) plots the $y$-averaged electron density $\langle N_e(x)\rangle_y \propto -\Delta\phi_e(x)$ at delay $\Delta t \sim 1$ ps, well before any hydrodynamic response. The curves for LP and CP reveal a plasma column wider than the Bessel beam focal profile $|J_0(k_\perp r)|^2$ in vacuum, where $k_\perp = k\sin\gamma$. Rings of electron density from ionization by the Bessel beam lobes are clearly seen, and are displaced from the rings of $|J_0(k_\perp r)|^2$. This is consistent with refraction of the later time slices of the Bessel beam pulse by plasma generated earlier in the pulse. This is confirmed by the Bessel beam propagation simulations discussed in Sec. III. Notable is the wider curve for LP. While hydrogen ionization saturates near beam center for both LP and CP, the larger peak electric field in the LP pulse generates more ionization in the wings.

**III. Simulations of femtosecond Bessel beam propagation, plasma generation, and heating**

Given the Bessel beam geometric focal length $L_{foc} \sim R/\tan\gamma$, generation of meter-length OFI plasma waveguides using beam radius $R < 3$ cm requires $\gamma < \sim 30$ mrad ($\sim 1.7°$). Such small approach angles make the Bessel beam susceptible to exclusion from the focal region when its generated electron density exceeds the effective critical density, $N_e > N_{cr,eff} = N_{cr} \sin^2\gamma$, where $N_{cr}$ and $N_{cr,eff}$ are the critical and effective critical densities. However, even for densities below $N_{cr,eff}$, the beam can be significantly refracted and distorted. For example, the plasma density $N_e \sim 10^{17}$ cm$^{-3}$, which is in the optimal range for multi-GeV acceleration by $\lambda \sim 800$ nm laser drivers (where $N_{cr} = 1.74 \times 10^{21}$ cm$^{-3}$) [10,11], is reached after hydrodynamic expansion from an initial density $N_e \sim 10^{18}$ cm$^{-3}$ [11,26,27]. For this density, Bessel beam exclusion will occur at at $\gamma < 1.4°$, but refraction and distortion can still be expected at larger $\gamma$. Furthermore, at the intensities required for plasma generation, a significant effect is nonlinear propagation of the pre-focus Bessel beam rays in the target gas.

These considerations imply that accurate simulation of OFI and heating requires self-consistent treatment of Bessel beam pulse propagation in the target gas. In earlier work, we perfomed self-consistent propagation simulations in the context of our 2-Bessel method for plasma waveguide generation [26]. In even earlier work, Bessel beam-plasma interaction had been self-consistently simulated for 100 ps pulses [43], where hydrodynamic evolution during the interaction was taken into acccount. In the present case of a 50 fs pulsed Bessel beam, there is ionization and but no hydrodynamic evolution during the pulse.

Here, we use our non-paraxial carrier-resolved electromagnetic propagation code YAPPE [27], based on the unidirectional pulse propagation equation [44], to establish the initial gas and plasma conditions for a subsequent hydrocode simulation (Sec. IV). We assume cylindrical symmetry, justified by the high fidelity Bessel beams we generate using phase correction [39]. YAPPE includes all important nonlinear propagation effects, including the nonlinear index of refraction of gas constituents [45], and ionization. The ionization model tracks the neutral and ion species $H_2$, $H_2^+$, $H$, and $H^+$ via coupled rate equations of atomic hydrogen, ionization and dissociation of molecular hydrogen (see Appendix B), and considers only the molecular vibrational ground states. The code calculates the ionization rate of molecular hydrogen using the MO-PPT model [46–49]. Following [49], we assume two-step ionization of molecular hydrogen: $H_2 \rightarrow H_2^+ + e$, $H_2^+ \rightarrow H + H^+ \rightarrow H^+ + H^+ + e$.

The simulation input pulse (40 mJ, 50 fs FWHM) is a $4^{th}$ order super-Gaussian profile with a $1/e$ field radius $w_0 = 1$ cm at its waist, giving $L_{foc} = 25$ cm for $\gamma = 2.3°$; the simulation Bessel beam peak vacuum intensity and profile are the same as in the experiment. The simulated pulse propagates from the axicon output through 30 cm of neutral $H_2$ at molecular density $1.6 \times 10^{18}$ cm$^{-3}$. The longitudinal average electron temperature after the pulse is calculated as $k_B T_e(r,z) = (2/3)[\int d\xi\, (dN_e/d\xi)]^{-1} \int d\xi\, (dN_e/d\xi)\, |\mathbf{p}|^2/2m$, where $\mathbf{p} = e\mathbf{A}(r,\xi;z)/c$ is the initial electron momentum at the position and time of ionization, $\mathbf{A}$ is the laser pulse vector potential, and the integration is taken over the full laser pulse envelope within the computation window.

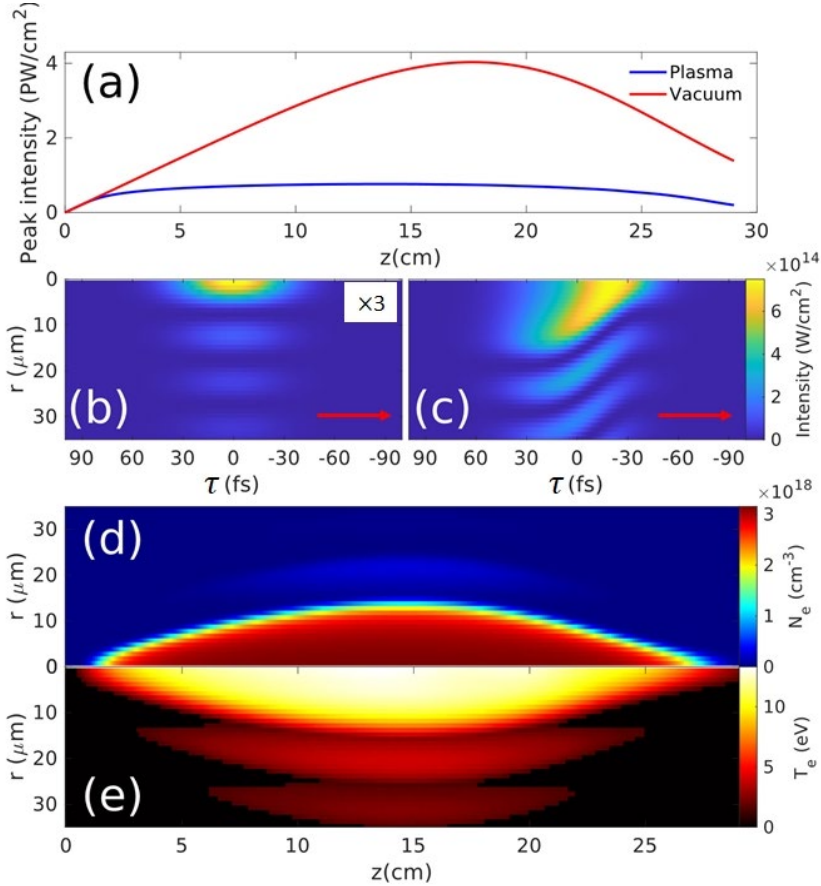

**Figure 3.** YAPPE simulation results for LP Bessel beam pulses (40 mJ, 50 fs FWHM, $4^{th}$ order super-Gaussian profile with $1/e$ field radius $w_0 = 1$ cm, $\gamma = 2.3°$). (a) Peak on-axis instantaneous laser intensity with (blue) and without (red) $H_2$ gas present. (b) Bessel beam intensity envelope ($\times 3$) at $z = 0.9$ cm. Red arrow indicates propagation direction. (c) Bessel beam intensity envelope at $z = 15$ cm. $\tau = t - z/v_g$ is the temporal coordinate in the moving window. (d) Electron density (e) Electron temperature

Figure 3 presents the YAPPE simulation results for LP, with Fig. 3(a) plotting the on-axis peak instantaneous (not cycle averaged) intensity vs. $z$ with and without 67 mbar $H_2$ gas present. The peak vacuum intensity of $4 \times 10^{15}$ W/cm$^2$ corresponds to our experimental conditions for a 50 fs, 150 mJ Bessel beam pulse. The bell-shaped profile along $z$ in vacuum is caused by the radial weighting of the incident beam profile on the axicon and its projection into the focal volume [21]. With the gas present, the profile flattens to a relatively constant peak instantaneous intensity $\sim 7 \times 10^{14}$ W/cm$^2$ owing to refraction of the Bessel beam from its self-generated plasma, and its central lobe (bounded by the first dark ring) expands from a radius of $\sim 7$ μm at $z = 0.9$ cm (Fig. 3(b)) to $\sim 16$ μm at $z = 15$ cm (Fig. 3(c)). Movies of the Bessel pulse propagation are available at [50].

The post-pulse electron density and temperature profiles are plotted in Figs. 3(d) and (e), where the peak values are $\sim 3 \times 10^{18}$ cm$^{-3}$ and $\sim 12$ eV. Bessel beam refraction by the plasma effectively improves the uniformity and efficiency of OFI heating by leveling the axial variation of peak intensity and expanding the OFI radius, particularly when the plasma density is close to or above $N_{cr,eff}$. The radial extent of the ionization and heating falls off at the ends of the plasma owing to the falloff in peak intensity there (blue curve in Fig. 3(a)). We note that at $\gamma = 2.3°$, $N_{cr,eff} \sim 3 \times 10^{18}$ cm$^{-3}$, consistent with the peak electron density plotted in Fig. 3(d).

A similar simulation was performed for CP, with all other laser and gas parameters the same. Figure 4(a) plots the peak instantaneous intensity with and without H$_2$ gas present, with refractive leveling of the peak intensity (here to $\sim 4 \times 10^{14}$ W/cm$^2$) similar to Fig. 3(a). The central lobe similarly widens between $z = 0.9$ cm and $z = 15$ cm (Fig. 4(b) and (c)) and peak electron density similarly reaches $\sim 3 \times 10^{18}$ cm$^{-3}$ (Fig. 4(d)) because the peak laser field, even for CP, is sufficient for saturated OFI of hydrogen. The main difference in the CP case is the higher peak electron temperature, which reaches $\sim 45$ eV (Fig. 4(e)).

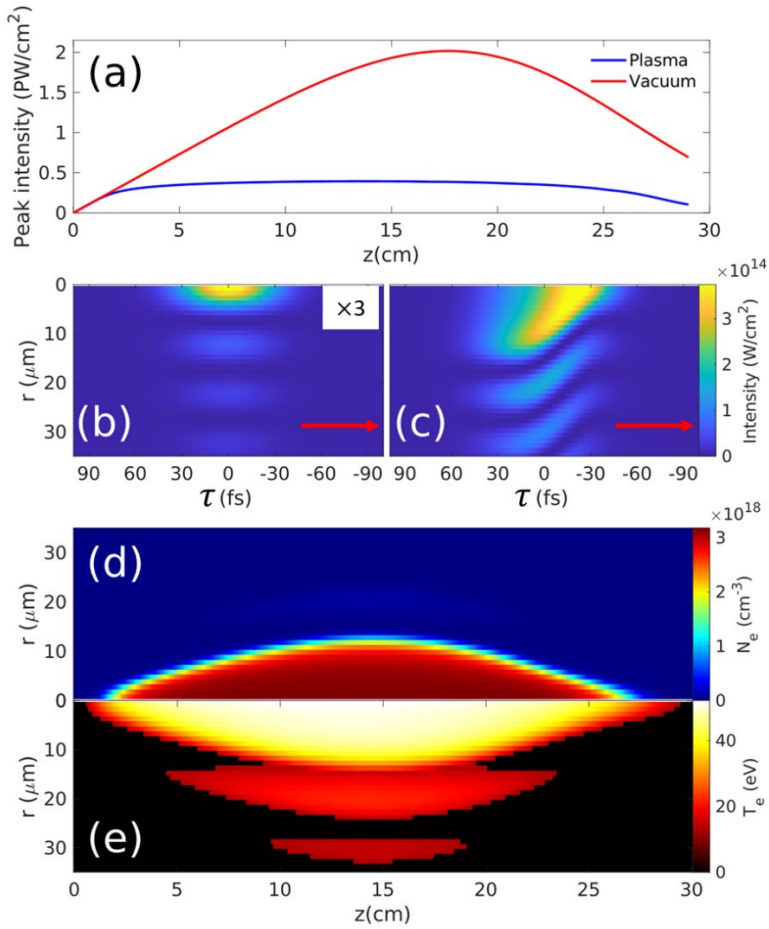

**Figure 4.** YAPPE simulation results for CP Bessel beam pulses (40 mJ, 50 fs FWHM, $4^{th}$ order super-Gaussian profile with $1/e$ field radius $w_0 = 1$ cm, $\gamma = 2.3°$). (a) Peak on-axis instantaneous laser intensity with (blue) and without (red) H$_2$ gas present. (b) Bessel beam intensity envelope ($\times 3$) at $z = 0.9$ cm. Red arrow indicates propagation direction. (c) Bessel beam intensity envelope at $z = 15$ cm. $\tau = t - z/v_g$ is the temporal coordinate in the moving window. (d) Electron density (e) Electron temperature.

## IV. Hydrodynamic simulations

We use the code SPARC [37] to simulate the hydrodynamic evolution up to 10 ns after the ultrashort pulse Bessel beam ionization and and heating of hydrogen gas. The code tracks the evolution of the velocity fields and the density and temperature profiles of the electrons and combined ion/neutral heavy particles (see Appendix C). It also tracks, via rate equations, the populations of $H_2(v = 1,2)$ (the first two hydrogen vibrational levels), $H_2^+$, H, and $H^+$. All simulations were performed in 1D cylindrical geometry, consistent with the symmetry of the Bessel beam and plasma generation in our experiments.

The initial pressure profile, which initiates the hydrodynamic evolution, was computed as the sum of the electron and ion pressures. The electron pressure is the product of the electron density and electron temperature profiles computed from the YAPPE simulations of Figs. 3 and 4. The ion pressure was initiated either at zero (room temperature ions) or at a level consistent with an ion temperature (3 eV) established by Coulomb explosion of $H_2^+$ during the Bessel beam interaction (see below). Figure 5(a) and 5(b) show the evolution of electron density (dashed curves) and combined heavy particle ($H_2, H_2^+, H, H^+$) density (solid curves) for LP and CP pulses. The number density of heavy particles, $N_{heavies}$, is the equivalent molecular density (or half the number of protons). The ultrafast heated plasma column drives a decreasing heavy particle density on axis and an expanding shock wave in the surrounding neutral gas. The electron density evolves as a widening and declining peaked profile with no concave guiding structure developing, in agreement with the measurements of Fig. 2. The profiles are consistent with the measurements of Fig. 2(a,a′) and 2(b,b′) and [27], with a relatively flat central electron density profile in the low $10^{17}$ cm$^{-3}$ range. One experimental feature not reproduced in the simulations is the transient off-centre peak in neutrals at $\Delta t \sim 4 - 5$ ns, a discrepancy likely related to modeling of multispecies shocks, and currently being studied. In our multi-GeV LWFA experiments [11], the full plasma waveguide is formed by self-waveguiding [27], where the central electron density profile forms the waveguide core and the self-waveguiding pulse ionizes the shock walls to form the cladding.

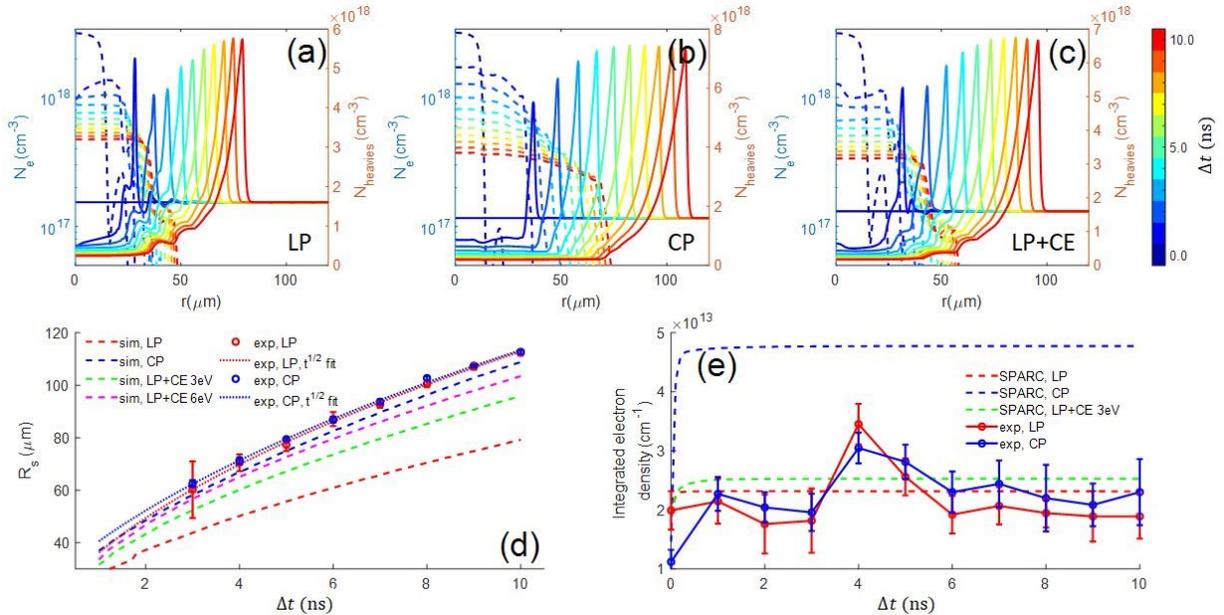

**Figure 5.** SPARC simulation result using YAPPE output for (a) LP and (b) CP as initial conditions. (c) LP+CE (linear polarization and Coulomb explosion-imposed ion temperature of $T_i = 3$ eV) (d) Simulated shock peak position $R_s$ vs time (from (a),(b), and (c)) and comparison with Figs. 2(a′), (b′). (e) Simulated electron density per unit length $Q(\Delta t) = \int dr 2\pi r N_e(r, \Delta t)$ for LP, CP, and LP+CE, overlaid on the LP and CP measurements from Fig. 2(c).

Quantitative comparison of experiments and simulations is presented in Fig. 5(d) and 5(e). Figure 5(d) plots the measured and simulated shock peak position $R_s$ vs time (from Figs. 2(a′)(b′)), with the $R_s \propto t^{1/2}$ cylindrical blast wave time evolution [41] fit to the experimental points. It is seen that the experimental curves for LP and CP are close together and line up with the CP simulation, while the LP simulation underesimates the shock position by up to ~40%. Figure 5(e) replots the electron density integral $Q(\Delta t)$ vs. $\Delta t$ from Fig. 2(c), showing reasonable agreement with the LP simulation, while the CP simulation overestimates the measurement by ~2 ×. The simulation plots show that plasma recombination is not significant over the first ~10 ns of hydrodynamic evolution.

The main discrepancy between experiment and simulation is that the LP and CP results are similar in the experiment but different in the simulations. In Fig. 5(b), the higher electron temperature in the CP case is seen to drive collisional ionization in the wings of the expanding electron profile near the shock; however, as seen in the measurements of Fig. 2(b), these lower density wings are not captured by interferometry. We therefore interpret the similar linear charge densities for LP and CP in Fig. 5(e) as reflecting the limited sensitivity of the interferometric measurements to low densities. Regarding the similarity of LP and CP experimental results in Fig. 5(d), we suggest that the simulation's neglect of the Coulomb explosion channel ($H_2^+ \rightarrow H^+ + H^+ + e$) underestimates the ion temperature in the LP case. Indeed, LP has a higher probability of driving Coulomb explosion via $H_2^+$ bond softening and ionization of intermediate vibrational states, with directly generated ion energies $\varepsilon_i$~4 eV [51,52] for approximately our intensity and pulsewidth. The ions experience a ~1 μm collisional mean free path and instantaneously boost the ion temperature to $T_i$~$(2/3)\varepsilon_i$~3 eV, contributing significantly to the subsequent hydrodynamic expansion. By contrast, the timescale for electron-ion energy transfer by the OFI-heated electrons at these densities is several nanoseconds, while cooling the electrons. We account for the Coulomb explosion boost to the ion temperature in an additional SPARC simulation for LP with the ion ($H^+$ and $H_2^+$) temperature initialized at $T_i = 3$ eV. The results are plotted in Fig. 5(c): while the electron density profiles are only slightly wider than in Fig. 5(a), the heavy particle profiles are closer to those of Fig. 5(b). This is also shown in Fig. 5(d), where the associated shock position ($R_s$) curve ("LP+CE 3 eV" in legend) is now in better agreement with experiment. It is seen that increasing initial ion temperature to $T_i = 6$ eV has only a minor effect ("LP+CE 6 eV" in legend). To summarize, while our intial expectation was that CP-ionized waveguides would expand faster than LP-ionized guides owing to greater electron heating (comparing Figs. 3(e) and 4(e)), our experimental results show only a small polarization effect. A likely explanaton is LP-favoured Coulomb explosions of $H_2^+$, which boosts the ion temperature enough to make LP competitive with CP.

We now use the results of the hydrocode simulation (corresponding to Fig. 5(c)) to compute the lowest order guided mode of the evolving plasma waveguide. We take as the waveguide plasma density profile the sum of the plasma and neutral atomic densities (accounting for plasma waveguide generation by self-waveguiding [27]), giving the plasma density profile vs. delay plotted in Fig. 6(a). Owing to the finite height and width of the shock wall-formed plasma waveguide cladding, all modes are quasibound to varying degree and are found using a Helmholtz solver to analyze the quasibound mode $k_\perp$ spectrum [26,27,38]. For low-leakage hydrodynamic plasma waveguides for application to LWFA, the lowest order quasibound mode is numerically indistinguishable from a bound eigenmode, here with a leakage attenuation length [38] more than an order of magnitude longer than a 1 m waveguide. The lowest order mode intensity profile of the hydrodynamically evolving waveguide is plotted in Fig. 6(b), showing tunability of the mode size as a function of delay, with the solid white line marking the evolution of the $1/e^2$ intensity spot radius $w_{ch}$. Noting the flat central region of electron density and its abrupt rise at the shock, the plasma waveguide can approximated as a step-index fiber [26], where the step index fiber parameter [53] is $V = ka(n_{core}^2 - n_{clad}^2)^{1/2} \approx kR_s(\Delta N_e/N_{cr})^{1/2}$, where $a \approx R_s$ is the waveguide core radius, $n_{core} = 1 - N_e^{core}/2N_{cr}$ and $n_{clad} = 1 - N_e^{clad}/2N_{cr}$ are the waveguide core and cladding refractive indices, and $\Delta N_e = N_e^{clad} - N_e^{core}$ is the plasma density difference between the core and cladding. For the conditions of Figs. 5 and 6, $V \sim 10$ so that the guided spot radius is [54] $w_{ch}^{step} = a(0.6484 + 1.619V^{-3/2} + \cdots) \approx 0.65R_s$, plotted as the dashed line in Fig. 6(b). The solid and dashed white curves agree well over 1 - 5 ns delay and diverge only slightly thereafter, indicating that the step index approximation is representative of the waveguide transverse profile and is reasonably accurate for estimates of the fundamental mode size.

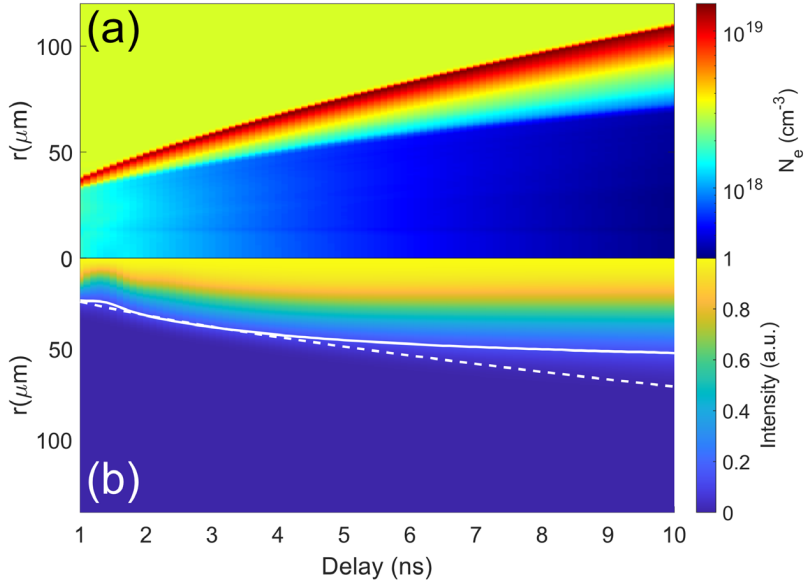

**Figure 6.** (a) Plasma waveguide profile evolution, obtained by adding the electron and neutral shock profiles of Fig. 5(b) (accommodating self-waveguiding). (b) Intensity profile of the lowest order quasi-bound mode vs. waveguide evolution. The profile at each delay is normalized. White solid line: $e^{-2}$ intensity mode radius. White dashed line: $w_{ch} \approx 0.65R_s$ (see text), based on [54].

## V. Simulations for meter length and longer plasma waveguides

To accelerate beyond 10 GeV, plasma waveguides must be longer than the $\leq 30$ cm guides we used to demonstrate multi-GeV LWFA [11,17,28]. Using modern multi-petawatt lasers [55] to drive acceleration to tens of GeV, plasma waveguides will need to support lowest order modes with $w_{ch} \sim 100$ μm and central plasma density $N_{e0} \sim 10^{16} - 10^{17}$ cm$^{-3}$, for which the 1D nonlinear dephasing length [56] is at least several meters. For example, recent simulations show that a LWFA driver pulsewidth 220 fs, energy 312 J, matched spotsize $w_0 = w_{ch} = 85$ μm, and a linearly ramped 6-m-long channel ($N_{e0}$ from $1.2 \times 10^{16}$ cm$^{-3}$ to $2.4 \times 10^{16}$ cm$^{-3}$), results in energy gain to 100 GeV [57]. These laser parameters are consistent with the ELI-Beamlines L4 laser [55].

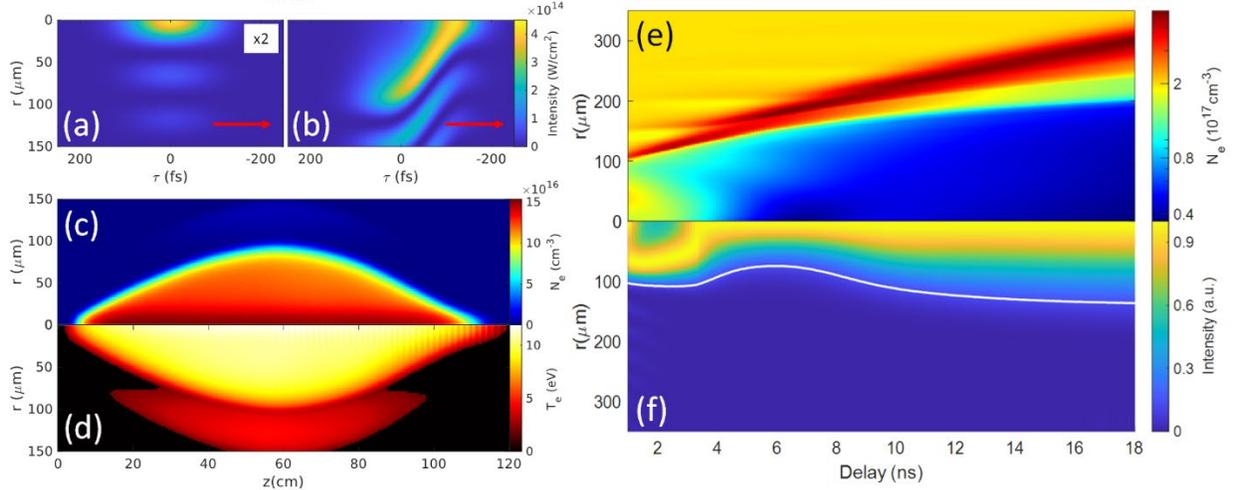

**Figure 7.** YAPPE and SPARC results for extendable meter-length plasma waveguide. (a) Bessel beam intensity envelope at $z = 4$ cm ($\times 2$). Red arrow is pulse propagation direction. (b) Bessel beam intensity envelope at $z = 60$ cm. (c) Electron density profile at $\Delta t \sim 0$. (d) Electron temperature profile at $\Delta t \sim 0$, indicating a peak of $\sim 11$ eV. (e) Evolution of plasma waveguide profile to $\Delta t = 18$ $ns$ after OFI by the Bessel beam, obtained by adding the electron and neutral shock profiles of Fig. 8 (accommodating self-waveguiding). (f) Evolution of the fundamental mode intensity profile, normalized at each delay. The white line indicates the mode's $e^{-2}$ intensity radius. *Simulation parameters*: Temporal grid size 0.12 fs, temporal window size 500 fs, average radial grid size 2 μm, radial window size 1.6 cm, Bessel beam axis approach angle $\gamma = 10$ mrad, input beam radius 1 cm, uniform hydrogen molecular density $1 \times 10^{17}$ cm$^{-3}$.

As an example of the typical waveguide generation requirements for this regime, Figs. 7 and 8 show simulations of the Bessel-beam-induced initial conditions in hydrogen gas and the evolution of its hydrodynamic response. Figure 7 shows results of YAPPE simulations of Bessel beam ionization and heating of uniform density hydrogen gas, here with a 0.5 J, 150 fs, λ=1.057μm LP Bessel beam pulse with $w_0 = 1$ cm and $\gamma = 10$ mrad axis approach angle, laser parameters typical of [55]. Here we simulate generation of a meter-long section of a plasma waveguide, for which longer guides would be concatenated with several possible schemes while scaling up the energy appropriately. Figures 7(a) and 7(b) show the Bessel beam spatiotemporal intensity profiles at $z = 4$ cm and $z = 60$ cm, illustrating the pulse distortion due to refraction from self-generated plasma, similar to Figs. 3 and 4—leading to a more axially uniform peak intensity (as in Figs. 3(a) and 4(a)) and a uniform on-axis plasma density (Fig. 7(c)). The electron temperature profile (Fig. 7(d)) peaks at ~11 eV.

The hydrodynamic response is modeled by SPARC [37] using the YAPPE-derived plasma density and electron temperature profiles at $z = 60$ cm (from Figs. 7(c) and 7(d)), along with an initial ion temperature of $T_i = 3$ eV from the hydrogen Coulomb explosion channel (see Sec. IV). The evolving profiles of electron and heavy particle density are plotted in Fig. 8. As before, we take the evolving plasma waveguide profile to be the sum of the plasma and neutral atomic densities, assuming that self-waveguiding has ionized the shock region. The result is plotted in Fig. 7(e) as function of delay $\Delta t$, with the associated fundamental quasi-bound mode profile evolution shown in Fig. 7(f), demonstrating delay-dependent tuning of the mode, with beam waists $w_{ch} >$ 100 μm. It is interesting to note that plasma generated by the outer lobes of the Bessel beam initiates multiple lower amplitude shock waves; by ~6 ns these shocks are swept up by the expanding plasma. This is likely due to the incomplete handling of multispecies shocks, and points to future hydrocode improvements.

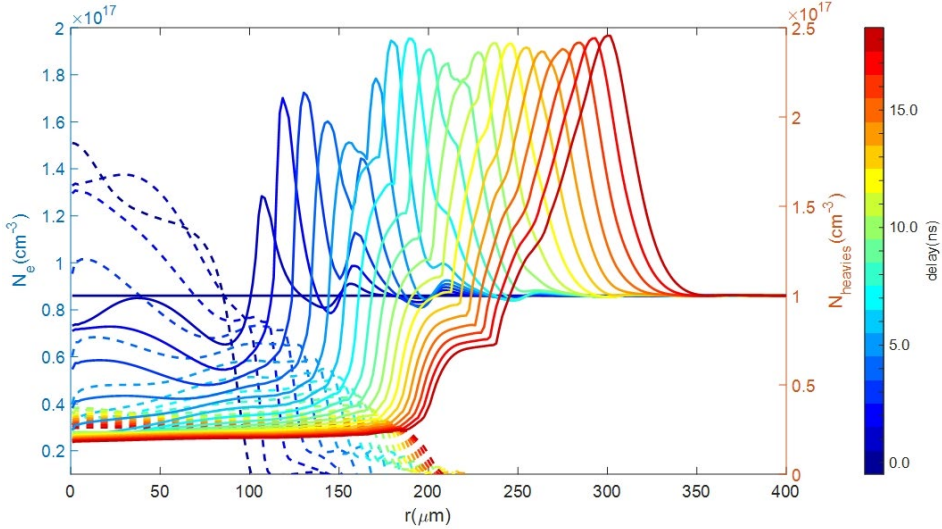

**Figure 8.** SPARC simulation of electron density and heavy particle profile evolution using YAPPE simulation output of Figs. 7(c) and 7(d) as initial conditions. For LP + CE (linear polarization and Coulomb explosion-imposed ion temperature of $T_i = 3$ eV)

## VI. Conclusions

We have presented an experimental and simulation study of hydrodynamically evolving plasma/gas channels initiated by optical field ionization and heating of hydrogen gas by short pulse Bessel beams. The initial electron density and temperature profiles that drive the simulated hydrodynamic evolution are generated by self-consistent propagation simulations of the short-pulse Bessel beam-gas interaction. This is essential for very shallow approach angles of the Bessel beam rays to the optical axis, required for meter-scale length plasmas. Under these conditions, self-generated plasma will refract and distort the Bessel beam, strongly affecting the density and temperature profiles ultimately generated. Using results of these simulations as initial conditions to a plasma hydrocode yields evolving electron density and neutral density profiles in good agreement with experiment, enabling realistic simulation of the guided mode structure. With improvement of model components such as reaction and ionization rates of hydrogen species, and treatment of multispecies shocks, the methods of this paper will be further refined for their continued use as a LWFA design tool.

**Acknowledgements**

The authors thank Scott Wilks and Brendan Reagan for useful discussions. This work was supported by the U.S. Department of Energy (DE-SC0015516, LaserNetUS
DE-SC0019076/FWP#SCW1668, and DE-SC0011375), the National Science Foundation (PHY2010511), and the Defense Advanced Research Projects Agency (DARPA) under the Muons for Science and Security Program. E. Rockafellow is supported by a NSF Graduate Research Fellowship (DGE 1840340).

## Appendix A: Details on two-color interferometry data and processing

The neutral gas and electron density profiles in Fig. 2 are extracted using two-color interferometry with λ=400 nm and λ=800 nm probe beams [27], with the probe geometry depicted in Fig. 1. The phase shift for each colour is $\Delta\phi^{(i)}(x,z) = k_i \int dy\, [(\Delta N_g(x,y,z)/N_{STP})\delta n_{g,STP}^{(i)} - N_e(x,y,z)/2N_{cr}^{(i)}] = \int dy\, (\delta\phi_g^{(i)} + \delta\phi_e^{(i)}) = \Delta\phi_g^{(i)} + \Delta\phi_e^{(i)}$ where $i = 1,2$ for the 800 nm and 400 nm probe beams, $\Delta N_g$ is the difference in hydrogen molecular density from uniform backfill gas density (at 67 mbar), $(1 + \delta n_{g,STP}^{(i)})$ is the refractive index of hydrogen at standard temperature and pressure (STP), $N_{STP}$ is hydrogen density at STP, and $\delta\phi_g^{(i)}$ and $\delta\phi_e^{(e)}$ are the neutral hydrogen and electron contributions to the refractive index. The neutral gas and plasma density profiles are then computed from Abel inversion [58] of the extracted phase shift profiles $\Delta\phi_g^{(i)}$ and $\Delta\phi_e^{(i)}$. The uncertainty of the density extraction was calculated according to [59] and limitations are discussed in [50]. The extracted electron and neutral density profiles for $\varepsilon_{Bessel} = 300$ mJ are plotted in Fig. A1.

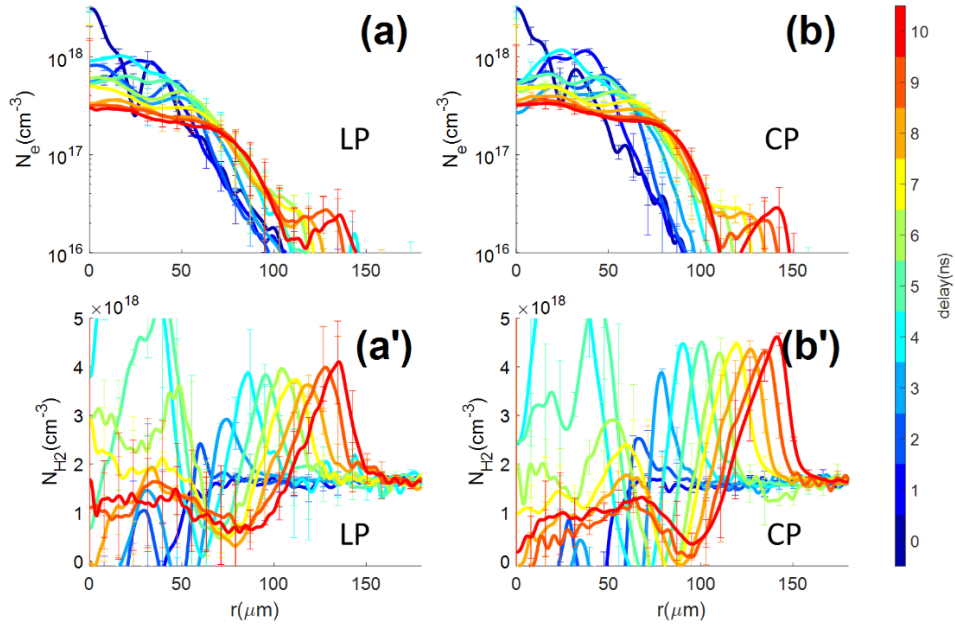

**Figure A1.** Evolution of electron density ((a),(a′)) and neutral density ((b),(b′)) profiles for LP and CP Bessel beam pulses with $\varepsilon_{Bessel} = 300$ mJ.

## Appendix B: Bessel beam propagation simulations

The simulations of beam propagation in this paper were performed using YAPPE (Yet Another Pulse Propagation Effort), an implementation of the unidirectional pulse propagation equation [44]. YAPPE numerically solves a system of ordinary differential equations (ODEs) of the form $\partial A_{k_\perp}(\omega,z)/\partial z = i2\pi Q_{k_\perp}(\omega) P_{k_\perp}(\omega,z) \exp(-i(k_z - \omega/v_g(\omega))z)$, where $A = A_{k_\perp}(\omega,z)$ is an auxiliary field related to the Fourier transform of the optical field by $E(\omega,z) = A\exp(ik_z z)$. The spectrum of radial spatial frequencies $k_\perp$ indexes a system of ordinary differential equations, which is solved using a GPU implementation of the MATLAB ODE45 function. The model is solved in a cylindrically symmetric geometry using discrete Hankel transforms. $P_{k_\perp}(\omega, z)$ is the nonlinear polarization of the medium including (for our simulations in this paper): third-order optical nonlinearities in $H_2$, ionization, and the plasma response, $\omega$ is angular frequency, $v_g(\omega)$ is the group velocity of the pulse as a function of frequency, $k_z = [(\omega/v_g(\omega))^2 - k_\perp^2]^{1/2}$ is the longitudinal spatial frequency, and $Q_{k_\perp}(\omega) = \omega/ck_z$.

For modeling Bessel beam nonlinear propagation, the hydrogen nonlinear refractive index $n_2$ [45] is scaled with the $H_2$ neutral density such that it goes to zero when all neutrals are ionized. The hydrogen ionization rate is computed with the MO-PPT model [46–49] assuming sequential ionization of $H_2$ molecules: molecular ionization ($H_2 \rightarrow H_2^+ + e$), followed by dissociation ($H_2^+ \rightarrow H + H^+$). The temporal resolution and domain sizes are $\delta t = 0.098$ fs and T = 200 fs. The average radial resolution is $\Delta r = 1.0$ μm and the radial domain size $r_{max} = 2.0$ cm. The axial step size is 100 μm. The moving window velocity is $v_w = c\cos(\gamma)$, consistent with the laser group velocity. Movies of the Bessel pulse propagation are available at [50].

**Appendix C: Hydrodynamic simulations using SPARC**

The SPARC code [37] solves the equations of multi-species, fully nonlinear gas dynamics (Euler equations). Here, the following species are tracked: e, $H^+$, $H_2^+$, H, $H_2$, and the first two vibration levels of $H_2(v = 1,2)$. Diffusive transport is accounted for in a separate, alternating step. Here, temperature, pressure, and transport coefficients are evolved in time using the ideal gas equation of state for each species. In a weakly excited gas, chemical kinetics is used to explicitly track energetic degrees of freedom such as vibrational states, electronic states, and charge states.

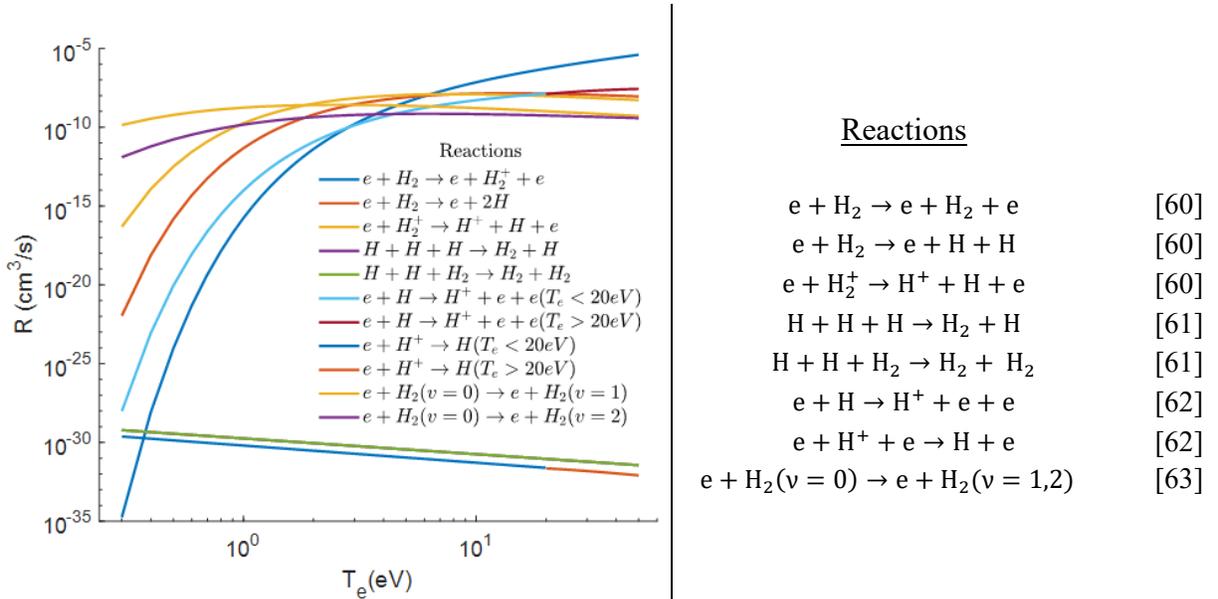

**Figure C1.** *Left side*: plots of electron collisional rates. *Right side*: associated reactions and citations.

Multi-species flow is handled as follows. Every species satisfies its own mass conservation equation. Species of similar mass are then gathered into groups with the same temperature (the kinetic degrees of freedom of intra-group particles are assumed to in thermal equilibrium at all times); all species within a group share a common velocity field. Here, all the heavy particles are in one equilibrium group ("heavies") and the light particles (electrons) are in the other group, each with its own temperature. Each such group satisfies its own momentum and energy equations, and couples to every other group via collision terms. Electron-ion collisions are modeled with Coulomb cross sections and the electron-neutral collisions are modeled with a fixed constant cross section $\sigma_{en} = 1.4 \times 10^{-15} \text{cm}^2$ [64]. Diffusion of one group with respect to another is tracked explicitly by means of collision terms.

The other reactions, plots of their rates, and the data source citations are presented in Fig. C1. The thermal conductivity and viscosity of the ions/electrons follow the Braginskii equations [65] and those of the neutrals follow the three-coefficient Sutherland's law [66]. All the heavy particles are in one equilibrium group, "heavies", with the same temperature, and the transport properties of the group are the density weighted sum of each species.

Corresponding to the simulation of Fig. 3, Fig. C2 plots the density profile evolution of several hydrogen species along with the evolution of the electron and heavy particle temperature and pressure profiles.

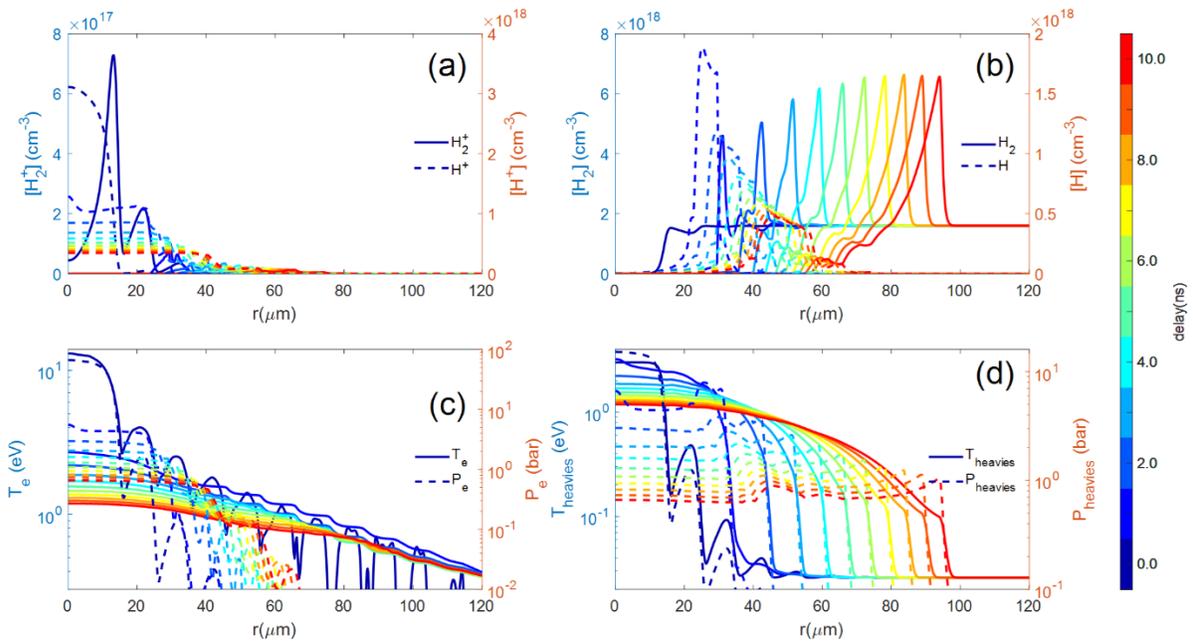

**Figure C2.** Density profile evolution for hydrogen species computed by SPARC using initial conditions of Fig. 3 plus initial ion temperature 3 eV. (a) Population density of $H_2^+$ and $H^+$. (b) Population density of $H_2$ and H. (c) Electron temperature and pressure (d) Ion temperature and pressure

# Supplementary Material: Benchmarking of hydrodynamic plasma waveguides for multi-GeV laser-driven electron acceleration

B. Miao[1], E. Rockafellow[1], J.E. Shrock[1], S.W. Hancock[1], D. Gordon[2] and H.M. Milchberg[1,3]

[1]Institute of Research in Electronics and Applied Physics and Dept. of Physics, University of Maryland, College Park, MD 20742
[2]Naval Research Laboratory, Washington DC 20375
[3]Dept. of Electrical and Computer Engineering, University of Maryland, College Park, MD 20742

## 1. Limitations on the neutral and electron density profile measurements

Two main factors affect the density profile measurement. The first is the spatial resolution of the imaging system, determined by the knife-edge method to be $\sigma_2 \approx$ 3 µm and $\sigma_1 \approx$ 6 µm for the 400-nm and 800-nm probes. We evaluate the effect of this limited resolution by simulating the plasma phase shift of a density profile in Fig. 5(a) at $t = 0$ ns. The ideal phase shift $\Delta\phi_{1,2}$ is calculated from the density profiles directly and the real phase shift is extracted by $\Delta\tilde{\phi}_i = Arg\left(\exp(i\phi_i) * h * \exp\left(-\left(\frac{x}{\sigma_i}\right)^2\right)\right)$, $i = 1,2$, where $h$ is the point spread function from the FFT-based phase extraction [1]. $\Delta\phi_{1,2}$ and $\Delta\tilde{\phi}_{1,2}$ are compared in Fig. S1(a), showing slight reduction in measured phase shift in the 800-nm interferogram. $\Delta\tilde{\phi}_{1,2}$ is converted to electron and neutral density profiles $\tilde{N}_e, \tilde{N}_{H2}$ and compared with the ideal density profiles $N_e, N_{H2}$ in Fig. S1(b). The on-axis electron density is lower by ~8%, in reasonable agreement but the neutral density profile is unphysical. Clearly, when the phase shift is dominated by plasma the neutral density cannot be extracted reliably due to the resolution effect.

The second factor is the ellipticity of the plasma, which in our case overestimates the electron density. In Fig. S1(c), the phase shift corresponding to Fig. 2(a) at $t = 0$ ns yields higher plasma density than full ionization when extracted individually. Informed by Fig. 3 and 4, we know the hydrogen molecules are fully ionized in the experiment. While this indicates the plasma cross section is non-circular, its exact shape is unknown. To simplify the scenario and produce physical results, we approximate the plasma cross section to be elliptical, with the long axis parallel to the probe beam path.

One hypothesis for plasma ellipticity is that the linearly polarized Bessel beam witnesses an approximately radial plasma density gradient. At each local azimuthal plane, the laser polarization varies from s to p and the plasma refractive index difference breaks the cylindrical symmetry. The detailed mechanism is outside the scope of this manuscript and will be studied in future experiments.

Therefore, the measured plasma phase shift should be corrected with an ellipticity factor $\epsilon$<1. The $\epsilon$ values are calculated at $t = 0$ ns for both probes, and $\epsilon_{800nm}$ is closer to 1 due to the lower optical resolution. The evolution of the ellipticity is determined empirically to be exponentially decaying to 1 at $t = \infty$. Fig. S2(d) plots the ellipticity model we used for the two-color probe beam, with a characteristic decay time of $\tau_\epsilon$=5 ns.

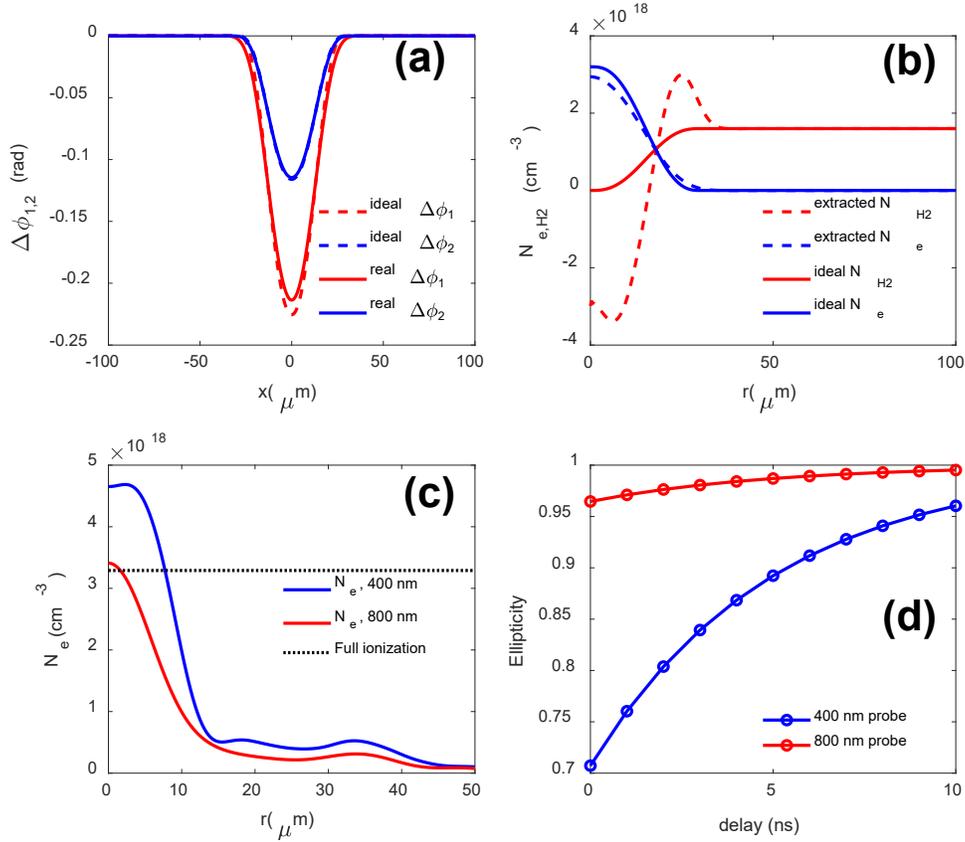

**Figure S1.** The effect of optical resolution and plasma ellipticity on density extraction. (a) Comparison between ideal (dashed) and simulated (solid) plasma phase shift of two-color probes ($\Delta\phi_{1,2}, \Delta\tilde{\phi}_{1,2}$). (b). Extracted (dashed) and ideal (solid) electron density profiles (blue) and neutral hydrogen (red). (c) Extracted electron density of Fig. 2(a) using the 400-nm (blue) and 800 (red) interferograms. The black dashed line labels the total electron density. (d) Ellipticity curve corresponding to the 400-nm (blue) and 800-nm (red) interferograms used in the extraction of Fig. 2(a).

## 2. Bessel beam propagation movies

Propagation animations are available at the following links. In each movie, the top panel plots the electron density profile, and the bottom panel plots the laser intensity.

**Table S1.** Description and links to the full propagation animations corresponding to Figs. 3, 4 and 7.

| Movie | Energy (mJ) | Axis approach angle (deg) | Polarization | Propagation length (cm) | Link |
|---|---|---|---|---|---|
| 1 | 40 | 2.3 | LP | 30 | Fig. 3 |
| 2 | 40 | 2.3 | CP | 30 | Fig. 4 |
| 3 | 500 | 0.57 | LP | 120 | Fig. 7 |